\documentclass[aps,reprint, superscriptaddress,preprintnumbers, amsmath,amssymb,amsfonts,nofootinbib]{revtex4-1}

\usepackage[hidelinks]{hyperref}
\usepackage{graphicx}
\usepackage[dvipsnames]{xcolor}
\usepackage{hyperref}
\usepackage{xspace}
\usepackage{ifdraft}
\usepackage{epstopdf}
\usepackage{slashed}
\usepackage{relsize}
\usepackage[normalem]{ulem}

\usepackage{tikz-feynman} 

\tikzfeynmanset{compat=1.1.0}
\tikzfeynmanset{/tikzfeynman/warn luatex = false}

\definecolor{jaxoblue}{HTML}{0086FF}

\newcommand{\bea}{\begin{eqnarray}}
\newcommand{\eea}{\end{eqnarray}}
\newcommand{\be}{\begin{equation}}
\newcommand{\ee}{\end{equation}}

\newcommand{\xvec}{\vec{x}}

\newcommand{\eqn}[1]{Eq.~(\ref{#1})}

\newcommand{\pd}{\partial}
\newcommand{\om}{\Omega_{\rm m}}

\newcommand{\tin}{t_{\rm in}}
\newcommand{\yvec}{\vec{y}}
\newcommand{\calom}{\mathcal{O}_m}

\definecolor{MyBlue}{rgb}{0.15,0.15,0.70}
\definecolor{linkblue}{rgb}{0,0,0.8}
\definecolor{linkgreen}{rgb}{0,0.5,0}
\hypersetup{
colorlinks=true,
citecolor=linkgreen,
linkcolor=linkblue,
urlcolor=linkblue
}

\newcommand{\andd}{\ , \quad \text{and}  \quad}

\newcommand{\appref}[1]{App.~\ref{#1}}
\newcommand{\figref}[1]{Fig.~\ref{#1}}

\newcommand{\xfl}{\xvec_{\rm fl}}

 \definecolor{MattOrange}{rgb}{1.0,0.4,0.2}

\begin{document}

\preprint{
}

\author{Yaniv Donath}
\affiliation{ Department of Applied Mathematics and Theoretical Physics,\\
University of Cambridge, Cambridge, CB3 OWA, UK}
\author{Matthew Lewandowski}
\affiliation{Institut f\"ur Theoretische Physik, ETH Z\"urich,
8093 Z\"urich, Switzerland}
\author{Leonardo Senatore}
\affiliation{Institut f\"ur Theoretische Physik, ETH Z\"urich,
8093 Z\"urich, Switzerland}

\title{Direct signatures of the formation time of galaxies}

\begin{abstract} 
We show that it is possible to directly measure the formation time of galaxies using large-scale structure.  In particular, we show that the large-scale distribution of galaxies is sensitive to whether galaxies form over a narrow period of time before their observed times, or are formed over a time scale on the order of the age of the Universe. Along the way, we derive simple recursion relations for the perturbative terms of the most general bias expansion for the galaxy density, thus fully extending the famous dark-matter recursion relations to generic tracers.

\end{abstract}

\maketitle

\section{Introduction and conclusions}

The establishment of the standard cosmological model, from the hot big bang at early times to the cosmological constant and cold dark-matter dominated late-time accelerated expansion, is one of the great triumphs of modern science.  It gives a depiction of a dynamical Universe that has evolved over billions of years from a dense cosmic soup to a sparse sprinkling of stars, galaxies, and dark-matter halos.  This familiar picture was not always obvious, however.  

For example, there was much debate in the second half of the twentieth century about the so-called 1948 steady-state model of the Universe \cite{Bondi:1948qk}.  This model proposed that properties of the Universe, including number and types of galaxies, did not change over time.  Empirical evidence, of course, eventually contradicted these ideas.  One such set of evidence was the observation that properties of galaxies, including color and estimated ages, changed with their measured redshifts (see for example \cite{stebbins, gamownote}), suggesting that the galaxies themselves evolved over time.  This confusion, though, is understandable.  {Indeed, we cannot watch objects in the Universe evolve for very long; we can only see static snapshots at various times in the past, making it quite challenging to {\it directly} probe cosmic time scales.}

{A concept that is related to, but distinct from, the time scale of cosmic evolution is what we call a cosmic \emph{response time}, i.e. the temporal extent to which the past influences galaxies at a given time.\footnote{{Mathematically, this is the time scale of support of the Green's function describing the response.}}  This in turn is related to the \emph{formation time} of galaxies, which is at least as long as the response time.}

In this work, we provide, as far as we can tell, the first directly cosmologically observable signals that are sensitive to the formation time of galaxies (or galaxy clusters and other gravitationally-bound objects in general).  {By studying the response time of galaxies,} we show that the static pictures that we take of the Universe (in galaxy surveys, for example) can contain unique signatures that are only possible if galaxies have been forming over time periods on the order of the age of the Universe. Even if we have an incredibly large amount of evidence that this must be the case, the possibility of a direct cosmological observation  is, to us, quite an extraordinary prospect.\footnote{{We stress that in this work, we are not concerned with ages or generic evolution of structures (for which there is abundant astrophysical evidence, some of which we mentioned above), but with the response time of structures.  Previous studies in this direction include numerical simulations and the so-called assembly bias \cite{Gao:2005ca, Croton:2006ys}, although it can be challenging to directly relate the latter to galaxy formation time \cite{Mao:2017aym}.} }

Furthermore, since our reasoning is based on the effective field theory of large-scale structure (EFT of LSS,~\cite{Baumann:2010tm, Carrasco:2012cv}), which is the unique theory of gravity, cold dark matter, baryons, and tracers on large scales, our conclusions do not depend on specific modeling choices about stars or galaxies.  Given the recent success of using the EFT of LSS to analyze galaxy clustering data {\cite{DAmico:2019fhj,Ivanov:2019pdj,Colas:2019ret,DAmico:2022gki, DAmico:2022osl}}, we now have the intriguing opportunity to explore the Universe in this exciting new way.

It has been known for some time (see e.g. \cite{coles2, Fry:1992vr}) that on large scales, the galaxy distribution can be expressed as a Taylor expansion in the fluctuations of the underlying dark-matter distribution, an approach that goes by the general name of the \emph{bias expansion} (for a modern review, see \cite{Desjacques:2016bnm}).  This makes intuitive sense, since galaxies tend to form in regions of space where the dark-matter density, and hence the gravitational potential, is highest.  In \cite{McDonald:2009dh} it was argued that this dependence should be on second spatial derivatives of the gravitational potential and gradients of the dark-matter velocity, and a straightforward extension allows for a dependence on spatial derivatives of these quantities. But is galaxy clustering only affected by the {nearby} dark-matter distribution at the time that we measure it (\emph{local in time}), or does the configuration of the dark matter at earlier times, {of order a Hubble time earlier,} have an impact (\emph{non-local in time})?  Said another way, given two identical {localized} dark-matter configurations at a given time, will the same galaxies always form, or do we need to know the whole history of that configuration?

This question was conceptually answered in \cite{Senatore:2014eva}, which pointed out that the most general dependence, based on the symmetries relevant to dark-matter and baryon dynamics and galaxy formation on large scales, which are the equivalence principle and diffeomorphism invariance (the non-relativistic limit of which is called Galilean invariance), is on second spatial derivatives of the gravitational potential, gradients of the matter velocity {(and the relative velocity directly)}, and their spatial gradients, integrated over all past times.  This makes the EFT of LSS generally local in space, but non-local in time.\footnote{{See also \cite{Carrasco:2013mua, Carroll:2013oxa} for discussions of non-local-in-time effects in dark-matter clustering.}} 

However, until now, the most advanced perturbative calculations have shown that the non-local-in-time bias expansion up to fourth order is mathematically equivalent to the local-in-time expansion  \cite{DAmico:2022ukl}.  As we show in this work, though, this is no longer true at fifth order, and thus it is possible to see distinctly non-local-in-time effects in the galaxy-clustering signal.  Measuring the size of these effects would then give us a direct indication of the formation time scale of galaxies.  {As a side observation, this time scale would also give a {\it direct} (versus indirect) lower bound on the age of the Universe. }

\paragraph*{Notes} We work in the Newtonian approximation where $\Phi ( \xvec , t)$ is the gravitational potential, $a(t)$ is the scale factor of the Universe, the Hubble parameter is defined by $H(t) \equiv \dot a(t) / a(t)$, and the overdot~`$\,\dot{}\,$' stands for a derivative with respect to $t$.   The dark-matter fluid is described by the overdensity $\delta(\xvec , t)$ and fluid velocity $\vec{v} ( \xvec , t)$.  The growth factor $D(t)$ is defined as the growing mode solution to the linear equation of motion for $\delta$, i.e. satisfies $\ddot D + 2 H \dot D - 3 \om H^2 D/ 2 = 0 $, where $\om (t )$ is the time-dependent matter fraction.  

The building blocks of Galilean scalars are the dimensionless tensors
\be \label{randpdefs}
r_{ij} \equiv \frac{2 \pd_i\pd_j \Phi}{3 \om a^2 H^2} \andd  p_{ij} \equiv -  \frac{D}{ a \dot D } \pd_i v^j  \ .
\ee
For brevity, we will always denote the traces $\delta^{ij} r_{ij}=\delta$ (which is true because of the Poisson equation) and $\delta^{ij} p_{ij} \equiv \theta$ (which is our definition of $\theta$). Then, for other contractions, we write the matrix products as simple multiplication, i.e. $r^2 = r_{ij}r_{ji}$, $r^2 p = r_{ij}r_{jk} p_{ki}$, $rprp = r_{ij} p_{jk} r_{kl} p_{li}$, and so on (repeated indices are always summed over).  We work in the so-called Einstein-de Sitter approximation, where the time dependence of perturbations is given by
\begin{align}
\begin{split} \label{edstimedep}
\delta^{(n)} ( \xvec , t) &  = \left( \frac{D(t)}{D(t')} \right)^n \delta^{(n)} ( \xvec , t')  \ , \\
 \theta^{(n)} ( \xvec , t) &  =  \left( \frac{D(t)}{D(t')} \right)^n  \theta^{(n)} ( \xvec , t') \ .
\end{split}
\end{align}
In this work, we focus on the lowest-derivative bias terms that are sufficient to establish our claims, and leave a discussion of higher-derivative bias (and counterterms) for future work.  Finally, we focus on the real space (as opposed to redshift space) prediction, which in any case is the leading signal {if one restricts observations to directions near the line of sight}.  We leave extending our results to redshift space to future work.  For a much more detailed explanation of the notation used here, see \cite{DAmico:2022ukl}.

%
\section{Complete bias expansion and recursion} 

We start by constructing the most general bias expansion for the galaxy overdensity $\delta_g  ( \xvec , t) \equiv (n_g ( \xvec , t) - \bar n_g (t) ) / \bar n_g(t)$, where $n_g ( \xvec , t)$ is the galaxy number-density field and $\bar n _g (t ) $ is the average number density of galaxies, that is consistent with the equivalence principle, diffeomorphism invariance, and is non-local in time.  Up to $N$-th order in perturbations, we have
\be
\delta_g ( \xvec , t) \big|_N = \sum_{n = 1}^N \delta_g^{(n)} ( \xvec , t)  \ , 
\ee
where the expression at $n$-th order is given by the non-local-in-time integral over the sum of all possible local-in-time functions $\mathcal{O}_m$ up to order $n$ \cite{Senatore:2014eva}
\begin{align}
\begin{split} \label{deltagnexpression}
\delta_g^{(n)} ( \xvec , t) & = \sum_{\mathcal{O}_m} \int^t dt'  H(t') c_{\mathcal{O}_m} ( t , t' ) \\
& \hspace{.7in} \times [ \mathcal{O}_m ( \xfl ( \xvec , t , t' ) , t')]^{(n)}  \ , 
\end{split}
\end{align}
evaluated along the fluid element
\be \label{fluiddef}
\xvec_{\rm fl} ( \xvec , t, t ' ) = \xvec + \int_{t }^{t'} \frac{ d t''}{a(t'') }  \vec{v} \left( \xvec_{\rm fl} ( \xvec , t  , t'' ) , t'' \right) \ ,
\ee
and we use the square brackets and superscript notation $[ \cdot ]^{(n)}$ to mean that we perturbatively expand the expression inside of the brackets and take the $n$-th order piece.\footnote{ {There was an interesting discussion \cite{Camelio:2015gda} as to whether intrinsic alignments {(see \cite{Vlah:2019byq} for an EFT description)} of galaxies are most affected by the gravitational field at late or early times {\cite{Catelan:2000vm, Hirata:2004gc, Schmitz:2018rfw}}.  Our non-local-in-time bias expansion \eqn{deltagnexpression} takes both possibilities into account.}}  Neglecting baryons, as they are a small effect \cite{Lewandowski:2014rca, Braganca:2020nhv}, in \eqn{deltagnexpression}, since $\delta_g$ is a Galilean scalar, the equivalence principle implies that the set of functions $\{ \mathcal{O}_m \}$ is given by all possible rotationally invariant contractions of the dark-matter fields $r_{ij}$ and $p_{ij}$, and integrating the $\mathcal{O}_m$ along the fluid element is the most general way to write a non-local-in-time expression for $\delta_g$.  All of the complicated details of galaxy-formation physics is then encoded in the functions $c_{\mathcal{O}_m}$, which are a priori unknown (from the EFT point of view) time-dependent kernels, {which physically can be thought of as the response of the galaxy overdensity to a given field at a given time}.  The local-in-time expansion is given by setting $  c_{\mathcal{O}_m} ( t , t') = c_{\mathcal{O}_m}(t)  \delta_D ( t - t')/H(t)$. {Notice that we do not include any time derivatives of $r_{ij}$ or $p_{ij}$ in the set $\{\mathcal{O}_m\}$ because these operators are not present in the strictly local-in-time limit (i.e. they would be suppressed with respect to other terms by $H/\omega_{\rm short} \ll 1$ where $1/\omega_{\rm short}$ is the time-scale of the relevant local-in-time physics) \cite{Senatore:2014eva}.  Thus, our expansion covers all Hubble-scale non-local-in-time effects.}   {From now on, in the list of functions $\{\mathcal{O}_m\}$,} we identify the subscript $m$ on $\mathcal{O}_m$ to denote that the function starts at order $m$, i.e. $m = 3$ for $\delta^2 \theta, \delta^3, r^2 p, \dots$.  

In this way, the bias expansion at order $n$ is reduced to an algorithmic procedure.  To create the list of seed functions $\{ \mathcal{O}_m \}$, we list all contractions up to $n$ factors of $r_{ij}$ and $p_{ij}$.  We then iteratively Taylor expand $\mathcal{O}_m ( \xfl ( \xvec , t , t'),t')$ around $\xvec$ using the recursive definition \eqn{fluiddef}, and take the $n$-th order piece.  After performing this expansion, we end up with an expression that can be cast in the following notation~\cite{DAmico:2022ukl}
\begin{align}
\begin{split} \label{omordern}
& [ \mathcal{O}_m ( \xfl ( \xvec , t , t' ) , t')]^{(n)}  = \\
& \hspace{.7in}  \sum_{\alpha = 1}^{n - m +1} \left( \frac{D(t')}{D(t)} \right)^{\alpha+m-1} \mathbb{C}_{\mathcal{O}_m, \alpha}^{(n)} ( \xvec , t )  \ .
\end{split}
\end{align}
The resulting bias functions $\mathbb{C}_{\mathcal{O}_m, \alpha}^{(n)} $, which we say are in the fluid expansion of the seed function $\mathcal{O}_m$, are \emph{defined} by the expansion in \eqn{omordern}, whose form is guaranteed by assuming the scaling time dependence of the dark-matter fields \eqn{edstimedep}, as well as the implied relation
\be \label{cscalingtime}
\mathbb{C}_{\mathcal{O}_m, \alpha}^{(n)} ( \xvec , t )  = \left( \frac{D(t)}{D(t')} \right)^{n} \mathbb{C}_{\mathcal{O}_m, \alpha}^{(n)} ( \xvec , t ' )   \ .
\ee
Plugging \eqn{omordern} into \eqn{deltagnexpression}, and defining the expansion coefficients 
\be \label{biasdefs}
c_{\mathcal{O}_m,\alpha} ( t ) \equiv \int^t dt' H(t') c_{\mathcal{O}_m} ( t , t' ) \left( \frac{D(t')}{D(t)} \right)^{\alpha+m-1 } \ ,
\ee
we finally have the most general expansion of the overdensity at order $n$ in terms of fields at the same time
\be \label{finalexpression}
\delta^{(n)}_g ( \xvec , t) = \sum_{\mathcal{O}_m} \sum_{\alpha = 1}^{n-m+1} c_{\mathcal{O}_m,\alpha}(t) \,\mathbb{C}^{(n)}_{\mathcal{O}_m,\alpha} ( \xvec , t) \ . 
\ee

\begin{figure}[t]
\includegraphics[width=0.45\textwidth]{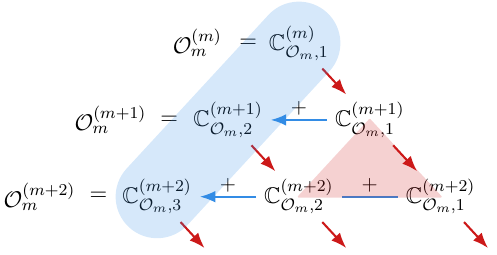}
 \caption{\footnotesize Diagrammatic representation of one way of using the recursion relations \eqn{completeness} and \eqn{fluidrecursion} to determine the full set of bias functions $\mathbb{C}^{(n)}_{\calom,\alpha}$ in the fluid expansion of a seed function $\mathcal{O}_{m}$.  The red arrows indicate the use of the fluid recursion \eqn{fluidrecursion}, while the blue arrows indicate the use of the completeness relation \eqn{completeness}.  Thus, the terms in the red shading ($\alpha < n - m + 1$) are determined by the fluid recursion \eqn{fluidrecursion} and the terms in the blue shading ($\alpha = n - m + 1$) are determined by the completeness relation \eqn{completeness}.   }
\label{recursion_diag}
\end{figure}

{There is in fact} a much simpler way to obtain the bias functions $\mathbb{C}_{\mathcal{O}_m, \alpha}^{(n)} $, using recursion relations, which is an additional key technical result of this work.  While the procedure described above is conceptually straightforward, it can be practically quite cumbersome (see the derivation at fourth order in \cite{DAmico:2022ukl}, for example).  The recursion relations come in two parts.  The first is the \emph{equal-time completeness relation}
\be \label{completeness}
\mathcal{O}_m^{(n)} ( \xvec , t )   =  \sum_{\alpha = 1}^{n - m +1} \mathbb{C}_{\mathcal{O}_m, \alpha}^{(n)} ( \xvec , t )  \ , 
\ee
which is trivially obtained by setting $t = t'$ in \eqn{omordern}, and where $\mathcal{O}_m^{(n)}$ is the standard expansion of $\mathcal{O}_m$ at $n$-th order in perturbations.  The second, which captures the consequences of expanding $\xfl$ in \eqn{omordern}, is the \emph{fluid recursion}
\begin{align}\label{fluidrecursion}
& \mathbb{C}_{\mathcal{O}_m,\alpha}^{(n)} ( \xvec , t) =  \\
&\frac{1}{n- \alpha - m+ 1}  \sum_{\ell = m+\alpha -1}^{n-1}  \partial_i \mathbb{C}^{(\ell)}_{\mathcal{O}_m , \alpha} ( \xvec , t )   \frac{\partial_i}{\partial^2} \theta^{(n - \ell)}  ( \xvec , t ) \ ,  \nonumber
\end{align}
which is valid for $n-\alpha - m + 1 > 0 $. {We explicitly derive \eqn{fluidrecursion} in \appref{fluidrecapp}.}  This recursion is reminiscent of the famous dark-matter recursion relations~\cite{Goroff:1986ep}, and provides, for the first time, a full generalization to generic biased tracers.   {We give a diagrammatic representation of this recursion relation in \figref{recursion_diag}.  }

{It is worth stressing that, unlike other treatments of biased tracers (such as \cite{Fry:1996fg, Tegmark:1998wm} and subsequent works), we do \emph{not} assume an instantaneous formation time of galaxies, nor do we assume a continuity equation for galaxies.  Indeed, \eqn{fluidrecursion} is a consequence of Galilean invariance (i.e. expanding $\xfl$), not of the conservation of galaxies.}

{Since we have formally done the integral over $t'$ in \eqn{biasdefs}, one might wonder where in \eqn{finalexpression} the non-local-in-time effect has gone.  Comparing \eqn{finalexpression} to the local-in-time expression
\be \label{finalexpressionloc}
\delta^{(n)}_{g, \text{loc}} ( \xvec , t) = \sum_{\mathcal{O}_m} c_{\mathcal{O}_m}(t) \,  \mathcal{O}_m^{(n)} ( \xvec , t )  \ , 
\ee
we see that the difference is in the basis functions of the expansion (which as we will discuss below control the possible functional forms of the clustering signals), since \eqn{finalexpression} is equivalent to \eqn{finalexpressionloc} under the restriction that, for all $\alpha$, $c_{\mathcal{O}_m,\alpha} (t ) = c_{\mathcal{O}_m} ( t )$. }


%
%
%

\section{Non-local-in-time bias in LSS}

We can {now} return to the main question posed by this work: \emph{is it possible to directly measure the effects of non-locality in time on galaxy clustering?}  In our perturbative description, this is equivalent to the the following mathematical question: \emph{does the basis for the non-local-in-time expansion \eqn{finalexpression} span a larger space than  the basis for the local-in-time expansion \eqn{finalexpressionloc}?}  The answer, as we will show below, is yes.  

As shown in \cite{DAmico:2022ukl}, the non-local-in-time and local-in-time expansions \emph{are} indeed equivalent up to fourth order in perturbations.\footnote{{Focusing on up to fourth order, \cite{Mirbabayi:2014zca, Desjacques:2016bnm} discussed how it is possible to map non-local-in-time terms into very special non-local-in-space terms.  The bases discussed there are degenerate with a local-in-time and local-in-space one, though \cite{DAmico:2022ukl}.}}  However, from the findings of this work, this seems to simply be a consequence of expanding to low orders in perturbation theory where there are too few independent spatially local and Galilean invariant functional forms available, since non-locality-in-time is generically expected in the bias expansion \cite{Senatore:2014eva}.  

{So, to discover a non-local-in-time effect, we look to fifth order.  In particular, we will now find the non-local-in-time basis for the expansion in \eqn{finalexpression}. }  To find the fifth-order functions {$\mathbb{C}^{(n)}_{\mathcal{O}_m,\alpha} $}, we form the set $\{ \calom\}$ by finding all rotationally invariant contractions of $r_{ij}$ and $p_{ij}$ up to fifth order.  Writing the first few terms, we have $\{\calom\} = \{ \delta, \theta , \delta^2, \delta \theta, \theta^2, r^2 , rp , p^2 , \dots \} $, and overall there are $63$ contractions with up to five factors.\footnote{{Here and in the rest of this work, since we work up to fifth order, we have already taken into account degeneracies that come from the fact that $r_{ij}^{(1)} = p_{ij}^{(1)}$ in terms that start at fifth order.  If we do not do this, there are {130} contractions with up to five factors.}}  We then find the functions $ \mathbb{C}_{\mathcal{O}_m, \alpha}^{(n)}$ for $n \leq 5$ either by expanding $\xfl$ as in \eqn{omordern}, or, equivalently, using the recursion relations \eqn{completeness} and \eqn{fluidrecursion}.   After this, there are $151$ bias functions for $n=5$.  However, as described in \appref{degenapp}, not all of these functions are independent.  In particular, we find a set of 122 degeneracy equations for $n = 5$, which means that there are $29$ independent functions that form the basis of the non-local-in-time expansion \eqn{finalexpression}.\footnote{{Using the Lagrangian basis expansion, \cite{Schmidt:2020ovm, Schmidt:2020tao} derived the number of independent fifth-order biases as 29, which is in agreement with our findings.}}  We provide all of the Fourier-space kernels relevant for the fifth-order expansion, and confirm all degeneracy equations, in an associated auxiliary file.

 %
 %

Next, we consider the basis of bias functions for the local-in-time expression \eqn{finalexpressionloc}.   At fifth order, this expansion starts with $63$ terms, however, as before, not all of them are linearly independent.  We find $37$ independent degeneracy equations, and hence $26$ independent functions for the local-in-time bias expansion at fifth order.  Indeed, this is three less than the non-local-in-time expansion, and hence \emph{the galaxy-clustering signal at fifth order is sensitive to whether or not galaxies form on time scales of order Hubble.}

We are now in a position to explicitly give the fifth-order basis derived for this work.
To be more concrete, we can write the fifth-order galaxy expansion in a basis with 26 elements that are local in time, and three that are non-local in time.  In this \emph{starting-from-time-locality (STL) basis}, we explicitly write 
  \be \label{lnlexpand}
 \delta_g^{(5)} (\xvec , t ) = \sum_{j=1}^{29}  \tilde b_j  ( t ) \mathbb{L}^{(5)}_j ( \xvec , t )  \ .
 \ee
 We choose the basis such that the elements with $j = 1 , \dots , 26$ are a basis of the local expansion \eqn{finalexpressionloc}.  Explicitly, we take $\mathbb{L}_j^{(5)} = \mathcal{O}_m^{(5)}$ with the corresponding $\mathcal{O}_m$ given by
 \begin{align}
 \begin{split}
 & \{  \delta , \theta , \delta \theta  , \theta^2  ,  r^2 ,  rp  ,  p^2   ,  \theta^3  ,   r^2p ,   rp^2 ,  p^3 , \\
 & \quad  r^2\theta ,   r p\theta , p^2 \theta ,  rp^3 ,  rprp , rp^2 \delta , r^3 \delta^2 , \\ 
& \quad   \delta^5 , r^3 \theta  , rp^2 \theta, rp \delta \theta , r^2 \theta^2 , rp \theta^2 , \delta \theta^3 , \theta^4   \} \ ,
 \end{split}
 \end{align}
 for $j = 1 , \dots , 26$.  Thus, the non-locality in time is contained in the final three basis elements, which we take to be 
 \begin{align}
 \begin{split}
 \mathbb{L}_{27}^{(5)}  = \mathbb{C}_{\delta,5}^{(5)}  \ , \quad   \mathbb{L}_{28}^{(5)}  =  \mathbb{C}_{r^2,4}^{(5)} \ , \quad   \mathbb{L}_{29}^{(5)} =    \mathbb{C}_{p^3,3}^{(5)}    \ . 
\end{split}
\end{align} 
Non-zero $\tilde b_{27}$, $\tilde b_{28}$, and $\tilde b_{29}$ can only come from non-local-in-time physics, so we call them non-local-in-time bias parameters.\footnote{{Here we reference the size of the physical bias parameters, which are generally made up of a combination of bare and counterterm contributions.}}  {We connect this basis to the so-called basis of descendants and show how fourth- and lower-order biases automatically consistently appear in \eqn{lnlexpand} in \appref{bodapp}.}

To see more quantitatively how the non-local-in-time bias parameters measure the time scale of galaxy formation, consider the expression \eqn{biasdefs} for the bias parameters.  Assuming that the kernel $c_{\mathcal{O}_m}(t , t') $ has support over a time scale of order $1 / \omega$ and expanding around the local-in-time limit, we have
\be
c_{\mathcal{O}_m , \alpha }  ( t ) \approx c_{\mathcal{O}_m} ( t ) \left( 1 + g_{\mathcal{O}_m,\alpha} (t ) {\frac{H}{\omega}} + \dots  \right) \ ,
\ee
where the $\dots$ represents terms higher order in $H/\omega$, and $g_{\mathcal{O}_m,\alpha} (t )\sim {\cal{O}}(1)$.  Since the non-local-in-time bias parameters $\tilde b_{27}$, $\tilde b_{28}$ and $\tilde b_{29}$ all vanish in the local-in-time limit, they are proportional to (at least)  $H/\omega$.  The size of the deviation from the first term, which is the local-in-time piece, is controlled by $H/\omega$: if there is a sizable deviation from the local-in-time limit, then $\omega \sim H$, and thus the time scale of the kernel $c_{\mathcal{O}_m}(t , t') $ is of the order $1/H$.\footnote{Of course, the measurement of a smaller deviation from the local-in-time limit means that the formation time scale could be correspondingly smaller.  {It could also mean that the theory is fine tuned in the sense that higher-order loop contributions accidentally largely cancel the lower-order biases.  On the other hand, it could also be that for a quasi-local-in-time theory, the coefficients of some non-local-in-time operators are accidentally large, which we refer to as being anomalous.  These accidents become more and more unlikely as one measures more parameters.}}  In our case, this happens if $\tilde b_{27}$, $\tilde b_{28}$, or $\tilde b_{29}$ are order unity.  This in turn would mean that the formation of the observed population of galaxies has been affected by the state of the Universe up to a Hubble time ago, and thus that it has formed on a time scale on the order of the age of the Universe.  

{It can be illuminating to momentarily consider a system that is truly local in time.  In this case, as we have discussed above, the bias parameters are expected to scale like $H / \omega_{\rm short} \ll 1$.  However, in the EFT, higher-order loops will generically contribute to the lower-order bias parameters.  Importantly, for a system that is truly local in time, those loops are expected to shift the bias parameters also by an amount that scales like $H  / \omega_{\rm short}$.  Given, though, that the cold dark-matter fluid is itself non-local in time \cite{Carrasco:2013mua, Carroll:2013oxa}, we expect that higher-order dark-matter loops will generically contribute $\sim \mathcal{O}(1)$ to the galaxy bias parameters.  {We remind the reader that by galaxies in this work, we mean gravitationally-bound structures that form around the non-linear scale at a given Hubble time.} }

%
%

\section{Observable signatures}

Until now, we have focused on the perturbative galaxy overdensity field itself.  In large-scale structure analyses, we typically compare to data using correlation functions (or $n$-point functions if they contain $n$ fields) of the overdensity fields of various tracers.  Thus, one way to measure the non-local-in-time effect that we have discovered in this work is in correlation functions.  Since we found that this effect arises at fifth order in perturbations, the lowest order observables sensitive to it are the two-loop two-point function through
\be \label{twolooppow}
\langle \delta_{g_1}^{(5)} ( \xvec_1 ) \delta_{g_2}^{(1)} ( \xvec_2 ) \rangle \ ,
\ee
the two-loop three-point function through
\be
\langle \delta_{g_1}^{(5)} ( \xvec_1 ) \delta_{g_2}^{(2)} ( \xvec_2 )  \delta_{g_3}^{(1)} ( \xvec_3 ) \rangle \ ,
\ee
the one-loop four-point function through
\be
\langle \delta_{g_1}^{(5)} ( \xvec_1 ) \delta_{g_2}^{(1)} ( \xvec_2 )  \delta_{g_3}^{(1)} ( \xvec_3 )  \delta_{g_4}^{(1)} ( \xvec_4 ) \rangle \ ,
\ee
the one-loop five-point function through
\be\label{fivepoint}
\langle \delta_{g_1}^{(5)} ( \xvec_1 ) \delta_{g_2}^{(2)} ( \xvec_2 )  \delta_{g_3}^{(1)} ( \xvec_3 )  \delta_{g_4}^{(1)} ( \xvec_4 )   \delta_{g_5}^{(1)} ( \xvec_5 )  \rangle \ ,
\ee
and the tree-level six-point function through
\be\label{sixpoint}
\langle \delta_{g_1}^{(5)} ( \xvec_1 ) \delta_{g_2}^{(1)} ( \xvec_2 )  \delta_{g_3}^{(1)} ( \xvec_3 )  \delta_{g_4}^{(1)} ( \xvec_4 )   \delta_{g_5}^{(1)} ( \xvec_5 ) \delta_{g_6}^{(1)} ( \xvec_6 )  \rangle \ ,
\ee
where we have used the subscript $g_i$ to denote possibly different tracer samples (each of which can have a different set of bias parameters), and we have taken all fields to be at the same time $t$ and dropped that argument to remove clutter.  

As two explicit examples, consider the contributions to the two-loop two-point function \eqn{twolooppow} and the tree-level six-point function \eqn{fivepoint} for $g_i = g$ for $ i = 1 , \dots, 6$.  Using the STL basis \eqn{lnlexpand}, we have the explicit non-local-in-time contributions  
\begin{align}
&\sum_{j=27}^{29} \tilde b_{j} \langle \mathbb{L}_j^{(5)} ( \xvec_1) \delta^{(1)}_g ( \xvec_2)  \rangle  \ , \\
&  \sum_{j=27}^{29} \tilde b_{j} \langle \mathbb{L}_j^{(5)} ( \xvec_1) \delta_{g}^{(1)} ( \xvec_2 )  \delta_{g}^{(1)} ( \xvec_3 )  \delta_{g}^{(1)} ( \xvec_4 )  \delta_{g}^{(1)} ( \xvec_5 )  \delta_{g}^{(1)} ( \xvec_6 )   \rangle \ , \nonumber
\end{align}
to the two-point and six-point functions respectively.
As we have seen, these would not be present in the galaxy correlation functions if galaxies formed in a local-in-time way.  {This makes them concrete, direct, observable signatures of the formation time of galaxies.}

\begin{acknowledgements}

We thank T. Abel and E. Komatsu for insightful comments on this manuscript.  Y.D. acknowledges support from the STFC.  L.S. is supported by the SNSF grant $200021\_213120$.

\end{acknowledgements}

\appendix

%
%
\section{Proof of fluid recursion}  \label{fluidrecapp}

{To derive \eqn{fluidrecursion}}, we will want to take $d / d t$ of \eqn{omordern}, which means that we will need to know $\partial_t \xfl ( \xvec , t , t')$.  To find that, we notice that by definition the fluid element satisfies the composition rule 
\be \label{comprule1}
\xfl \left( \xfl ( \xvec , t_{\rm in} , t ) , t , t '  \right) = \xfl ( \xvec , t_{\rm in} , t '  ) \ .
\ee
Since the right-hand side is independent of $t$, this implies
\be
\frac{d}{dt} \xfl \left( \xfl ( \xvec , t_{\rm in} , t ) , t , t '  \right)  = 0 \ .
\ee
Using the chain rule, and 
\be \label{xfldiffeq}
\frac{d}{dt }  \xvec_{\rm fl} ( \xvec , t_{\rm in} , t)  =\frac{1}{a(t)}  \vec{v}  ( \xfl( \xvec , t_{\rm in} , t), t) \ ,
\ee
which follows immediately from the definition of $\xfl $ \eqn{fluiddef}, this implies
\begin{align}
\begin{split}
0 & = \Big[ \frac{\partial}{\partial t } \xfl ( \yvec , t , t' ) + \\ 
& \hspace{.4in} \frac{v^i ( \yvec , t ) }{a(t)} \frac{\partial}{ \partial y^i} \xfl ( \yvec , t , t' )  \Big]\Big|_{\yvec = \xfl ( \xvec , \tin , t )}  \ .
\end{split}
\end{align}
Since the initial $\tin$ is arbitrary, we can take $\tin = t$, which gives
\be \label{otherxflderiv}
\left( \frac{\partial}{\partial t } + \frac{v^i ( \xvec , t ) }{a(t)} \frac{\partial}{\partial x^i} \right) \xfl ( \xvec , t , t' ) = 0 \ . 
\ee
 This equation simply says that the convective derivative of the fluid element is zero, which makes intuitive sense since the convective derivative is defined to be along the fluid flow.  

Now we take $d / dt$ of both sides of \eqn{omordern}.  The right-hand side is simple, and we have (defining ${D^\alpha_m(t' ,t )} \equiv (D ( t' ) / D(t) )^{\alpha + m  -1}$ to reduce clutter)
\begin{align} \label{dbydt2}
& \frac{D(t)}{\dot D(t)}  \frac{d}{dt} \sum_{\alpha = 1}^{n - m +1} D^\alpha_m(t',t)  \mathbb{C}_{\mathcal{O}_m, \alpha}^{(n)} ( \xvec , t )  =  \\
&\hspace{.2in}      \sum_{\alpha = 1}^{n - m +1} D^\alpha_m(t',t )  (n -\alpha - m  + 1)  \mathbb{C}_{\mathcal{O}_m, \alpha}^{(n)} ( \xvec , t )  \ , \nonumber
\end{align}
where we have used \eqn{cscalingtime} for the time dependence of $ \mathbb{C}_{\mathcal{O}_m, \alpha}^{(n)} $.  

On the left-hand side, we have
\begin{align}
& \frac{d}{dt} [ \mathcal{O}_m ( \xfl ( \xvec , t , t' ) , t')]^{(n)}   =   \left[ \frac{d}{dt}  \mathcal{O}_m ( \xfl ( \xvec , t , t' ) , t') \right]^{(n)} \nonumber \\
& = \left[ \frac{\partial}{\partial t} x_{\rm fl}^i ( \xvec , t , t' ) \frac{\partial}{\partial y^i}  \mathcal{O}_m ( \yvec , t') \Big|_{\yvec = \xfl ( \xvec , t , t') } \right]^{(n)} \nonumber \\
& = \left[ - \frac{v^j ( \xvec , t )}{a(t)}  \frac{\partial}{\partial x^j} x_{\rm fl}^{i} ( \xvec , t , t')  \frac{\partial}{\partial y^i}  \mathcal{O}_m ( \yvec , t') \Big|_{\yvec = \xfl ( \xvec , t , t') } \right]^{(n)} \nonumber \\
& = \left[ - \frac{v^j ( \xvec , t )}{a(t)}  \frac{\partial}{\partial x^i}  \mathcal{O}_m ( \xfl ( \xvec , t , t')  , t') \right]^{(n)}   \label{dtomexpand} \\
& = \frac{\dot D(t)}{D(t)} \left[ \frac{\partial_i}{\partial^2} \theta  ( \xvec , t ) \frac{\partial}{\partial x^i}  \mathcal{O}_m ( \xfl ( \xvec , t , t')  , t') \right]^{(n)}  \nonumber \\ 
& = \frac{\dot D(t)}{D(t)}  \sum_{\ell = m}^{n-1}   \frac{\partial_i}{\partial^2} \theta^{(n-\ell)}  ( \xvec , t ) \frac{\partial}{\partial x^i}  \left[  \mathcal{O}_m ( \xfl ( \xvec , t , t')  , t') \right]^{(\ell)} \ ,  \nonumber
\end{align}
where we have used \eqn{otherxflderiv} to go from the second to third line, the chain rule to go from the third to fourth line, and the definition of $\theta$ from \eqn{randpdefs} in the fifth line.  Now, we use \eqn{omordern} to replace $\left[  \mathcal{O}_m ( \xfl ( \xvec , t , t')  , t') \right]^{(\ell)} $ to get
\begin{align} \label{dbydt1}
 & \hspace{-.1in}  \frac{D(t)}{\dot D(t)}  \frac{d}{dt} [ \mathcal{O}_m ( \xfl ( \xvec , t , t' ) , t')]^{(n)} \\
& = \sum_{\ell = m}^{n-1}   \sum_{\alpha = 1}^{\ell - m + 1} D^\alpha_m(t' ,t )  \frac{\partial_i}{\partial^2} \theta^{(n - \ell)}  ( \xvec , t )  \partial_i \mathbb{C}^{(\ell)}_{\mathcal{O}_m , \alpha} ( \xvec , t )  \nonumber \\ 
& =   \sum_{\alpha = 1}^{n - m } D^\alpha_m(t' ,t )   \sum_{\ell = m+\alpha -1}^{n-1}  \frac{\partial_i}{\partial^2} \theta^{(n - \ell)}  ( \xvec , t )  \partial_i \mathbb{C}^{(\ell)}_{\mathcal{O}_m , \alpha} ( \xvec , t ) \nonumber
\end{align}
where we have simply changed the order of the sums between the second and third lines.  Equating the coefficients of each power of $D(t')$ in \eqn{dbydt2} and \eqn{dbydt1} then gives our recursion relation \eqn{fluidrecursion}.  

%
%
\section{Degeneracy equations}  \label{degenapp}

 As mentioned {in the main text}, not all of {the bias functions $\mathbb{C}^{(n)}_{\mathcal{O}_m,\alpha}$ at a given $n$} are linearly independent in the sense that
\be \label{degeneq}
\sum_{\mathcal{O}_m} \sum_{\alpha = 1}^{n-m+1}  d^{(n)}_{i, \mathcal{O}_m,\alpha  } \mathbb{C}^{(n)}_{\mathcal{O}_m,\alpha} ( \xvec , t) = 0 \ , 
\ee
 for some time-independent coefficients $d^{(n)}_{i, \mathcal{O}_m,\alpha  }$ for $i = 1 , \dots , N_d^{(n)}$, where $N_d^{(n)} \equiv \text{rank}[ d^{(n)} ] $ is the number of independent degeneracy equations.  In particular, we find $N_d^{(5)}=122$, and \cite{DAmico:2022ukl} found $N_d^{(4)}=73$. {Additionally, letting $N_{\mathbb{C}}^{(n)}$ be the number of $\mathbb{C}^{(n)}_{\mathcal{O}_m,\alpha}$ functions that result after the procedure described in the main article, we find $N_{\mathbb{C}}^{(5)} = 151$ and \cite{DAmico:2022ukl} found $N_{\mathbb{C}}^{(4)} = 88$.  }  Finally, using $N_b^{(n)} \equiv N_{\mathbb{C}}^{(n)} - N_{d}^{(n)}$ to denote the number of basis elements at order $n$, this means that $N_b^{(5)} = 29$ and $N_b^{(4)}=15$.\footnote{For completeness, we also have $N_b^{(3)}=7$, $N_b^{(2)}=3$, and $N_b^{(1)}=1$ with this method \cite{Angulo:2015eqa}.}  We confirm all of the fifth-order degeneracy equations in the associated ancillary file.  
 
 Thus, one can solve the degeneracy equations \eqn{degeneq} in terms of $N_b^{(n)}$ basis elements, which we denote generically as $\mathbb{E}^{(n)}_j ( \xvec , t)$ for $j = 1 , \dots , N_b^{(n)}$. Since this is a basis, all of the original functions can be written in terms of it, so we have
 \be \label{degensol}
 \mathbb{C}^{(n)}_{\mathcal{O}_m,\alpha} ( \xvec , t)  = \sum_{j=1}^{N_b^{(n)}} A^{(n)}_{\mathcal{O}_m,\alpha, j} \, \mathbb{E}^{(n)}_j ( \xvec , t)  \ ,
 \ee
 for some time-independent coefficients $A^{(n)}_{\mathcal{O}_m,\alpha, j}$.   Plugging \eqn{degensol} into \eqn{finalexpression} then gives
 \be \label{generalbasisexpand}
 \delta_g^{(n)} (\xvec , t ) = \sum_{j=1}^{N_b^{(n)}}   e^{(n)}_j  ( t ) \mathbb{E}^{(n)}_j ( \xvec , t )  \ ,
 \ee
 where $e^{(n)}_j( t ) = \sum_{\mathcal{O}_m} \sum_{\alpha = 1}^{n-m+1} c_{\mathcal{O}_m,\alpha} (t) A^{(n)}_{\mathcal{O}_m,\alpha,j}$.  The coefficients $e_j (t )$ are called \emph{bias parameters}, and we have now written the galaxy overdensity in terms of the minimal number of linearly independent functions.

%
%
%
\section{Basis of descendants}  \label{bodapp}

 Another, perhaps more natural, choice of basis functions is the so-called basis of descendants \cite{Angulo:2015eqa}, where if $\mathbb{C}_{\mathcal{O}_m,\alpha}^{(n)}$ is used at order $n$, then $\mathbb{C}_{\mathcal{O}_m,\alpha+1}^{(n+1)}$ is used at order $n+1$.  We write the fifth-order expansion in the basis of descendants as
  \be \label{bodexpand}
 \delta_g^{(5)} (\xvec , t ) = \sum_{j=1}^{29}   b_j  ( t ) \mathbb{B}^{(5)}_j ( \xvec , t )  \ .
 \ee

{As shown below, the first $15$ terms in \eqn{bodexpand} are determined by the fourth-order terms.  That is, for $j = 1 , \dots, 15$, the $b_j$ in \eqn{bodexpand} are the same as those in~\cite{DAmico:2022ukl}, and the basis functions are given by
 \be
  \mathbb{B}_j^{(5)} =   \mathbb{B}_j^{(4)} \Big|_{\mathbb{C}_{\mathcal{O}_m,\alpha}^{(4)} \rightarrow \mathbb{C}_{\mathcal{O}_m,\alpha}^{(5)} }  
 \ee
where the  $\mathbb{B}_j^{(4)}$ are given explicitly in \cite{DAmico:2022ukl}. } For the new elements derived here, i.e. $j= 16 , \dots , 29$, we have $\mathbb{B}_j^{(5)} = \mathbb{C}_{\mathcal{O}_m,\alpha}^{(5)} $, where the indices $\mathcal{O}_m,\alpha$ take the following values for the given $j$
\be
\normalsize
  \begin{tabular}{|c|c|c|c|c|c|c|c|} 
  \hline
  $j:$ 	 	&	 16	&  17 	&  18 & 19 & 20 & 21 & 22  \\ \hline 
  $\mathcal{O}_m,\alpha:$		&	$\delta , 5$ & $\delta^2 , 4$ & $r^2,4$  & $\delta^3 , 3$  & $ r^3 , 3$ & $r^2 \delta , 3 $ & $\delta^4 , 2 $  \\
  \hline \hline
  
\hline
  $j:$ 	 	&	 23	&  24 	&  25 & 26 & 27 & 28 & 29  \\ \hline 
  $\mathcal{O}_m,\alpha:$		&	$r^3 \delta,2$ & $r^4,2$ & $\delta^5,1$  & $r^5,1$  & $r^4 \delta,1$ & $r^3 \delta^2,1$ & $p^3,3$  \\
  \hline
   \end{tabular} \ . 
\ee
We also note that fifth order is the first time that $\partial_i v^j$ has to be used as a seed function to form a basis, for example through $\mathbb{C}^{(5)}_{p^3 , 3}$ above.  This is contrasted with the case at fourth order where $\partial_i \partial_j \Phi$ is enough \cite{DAmico:2022ukl}.   

Converting between the STL basis and the basis of descendants, we find the following expression for the non-local-in-time bias parameters and the basis-of-descendants bias parameters
\begin{align}
\begin{split}
& \tilde b_{27} = b_1 - 4 b_2 + 6 b_3 - 4 b_4 + 90 b_8 - 76 b_9 + b_{16}  \ , \\
& \tilde b_{28}= b_{18} - b_{9}   \ , \\ 
& \tilde b_{29} =  - \frac{4 b_8}{3} + \frac{4 b_9 }{3} - \frac{10 b_{11}}{3} + \frac{7 b_{20}}{3} + b_{29}  \ . 
\end{split}
\end{align}

%
%

\section{Lower-order bias parameters} 

 Here we show how bias parameters at {fourth order} appear automatically as biases {at fifth order}.  {For notational convenience, in this Appendix we will use $\Gamma$ as the combined index $\calom, \alpha$, as in $\mathbb{C}_{\Gamma}^{(n)} \equiv \mathbb{C}_{\calom, \alpha}^{(n)}$, and $\Gamma_n$ as the set of the relevant $\mathcal{O}_m$ and $\alpha$ at order $n$, as defined in the sum in \eqn{finalexpression}.}  We start with the fifth-order degeneracy equations.  It turns out, as we explicitly check in the ancillary file, that the full set of degeneracy equations satisfied by $\mathbb{C}^{(5)}_{\Gamma} $, \eqn{degeneq} with $n = 5$, {can be put in} the block form
\begin{align}
\begin{split} \label{npl1degens}
0  =\sum_{\Gamma \in \Gamma_{4}}  d^{(4)}_{i,\Gamma } \mathbb{C}^{(5)}_{\Gamma} ( \xvec , t)+\sum_{\Gamma \in \Gamma_{5}\setminus\Gamma_{4}}  \tilde d^{(5)}_{i,\Gamma } \mathbb{C}^{(5)}_{\Gamma} ( \xvec , t)  \  ,
\end{split}
\end{align}
for $i = 1 , \dots , N_d^{(5)}$, with $d^{(4)}_{i,\Gamma }=0$ for $i\in [ N_d^{(4)}+1,N_d^{(5)} ] $ and $\tilde d^{(5)}_{i,\Gamma }=0$ for $i\in [ 1,N_d^{(4)} ] $.  {For $i = 1 , \dots , N_d^{(4)}$,  the second term on the right-hand side of \eqn{npl1degens} vanishes, so the $\mathbb{C}^{(5)}_{\Gamma}$ with $\Gamma \in \Gamma_{4}$ satisfy the same equations as the fourth-order functions, \eqn{degeneq} with $n = 4$.}  Therefore we can write them in an analogous way to the $n = 4$ case of \eqn{degensol}, that is
\be \label{degensolnp1}
 \mathbb{C}^{(5)}_{\Gamma} ( \xvec , t)  = \sum_{j=1}^{N_b^{(4)}} A^{(4)}_{\Gamma, j} \, \mathbb{E}^{(5)}_j ( \xvec , t)  \ ,
 \ee
for $\Gamma \in \Gamma_4$, with 
  \be \label{e5def}
\mathbb{E}_j^{(5)}  \equiv \mathbb{E}_j^{(4)}  \Big|_{\mathbb{C}_{\Gamma}^{(4)} \rightarrow \mathbb{C}_{\Gamma}^{(5)} }   \ ,
\ee
for $j = 1 , \dots , N_b^{(4)}$.  Said another way, since the $\mathbb{E}_j^{(4)}$ are just linear combinations of some $\mathbb{C}_{\Gamma}^{(4)}$, we define $\mathbb{E}_j^{(5)}$ for $j = 1, \dots, N_b^{(4)}$ to be the same expressions as $\mathbb{E}_j^{(4)}$, but with $\mathbb{C}_{\Gamma}^{(4)}$ replaced with $\mathbb{C}_{\Gamma}^{(5)}$, i.e.
\begin{align}
  \begin{split}
&  \mathbb{E}_j^{(4)} (\xvec , t) = \sum_{\Gamma \in \Gamma_4} \beta^{(4)}_{j , \Gamma} \mathbb{C}_{\Gamma}^{(4)} ( \xvec , t) \ ,\\
& \mathbb{E}_j^{(5)} (\xvec , t) = \sum_{\Gamma \in \Gamma_4} \beta^{(4)}_{j , \Gamma} \mathbb{C}_{\Gamma}^{(5)} ( \xvec , t) \ ,
\end{split}
\end{align}
for some coefficients $\beta^{(4)}_{j , \Gamma}$.  

Now, the bias expansion at fifth order is
 \be \label{npl1biasexpand}
  \delta_g^{(5)} (\xvec , t )  = \sum_{\Gamma \in \Gamma_{5}} c_{\Gamma} ( t ) \mathbb{C}_{\Gamma}^{(5)} ( \xvec , t ) \ . 
 \ee
 The sum above can be split into a sum over $\Gamma \in \Gamma_4$ and a sum over $\Gamma \in \Gamma_{5} \setminus \Gamma_4$.   For the sum over $\Gamma_4$, we have 
 \be \label{lowerbiaseq}
 \sum_{\Gamma \in \Gamma_{4}} c_{\Gamma} ( t ) \mathbb{C}_{\Gamma}^{(5)} ( \xvec , t ) = \sum_{j=1}^{N_b^{(4)}}   e^{(4)}_j  ( t ) \mathbb{E}^{(5)}_j ( \xvec , t )  \ , 
 \ee
where we have used \eqn{degensolnp1} and the definition of $e_j^{(4)}(t)$ below \eqn{generalbasisexpand}.  Thus, the degeneracy equations \eqn{npl1degens} ensure that it is exactly the fourth-order bias parameters $e_j^{(4)} (t)$ that appear in \eqn{lowerbiaseq}.  Then, for the sum over $\Gamma \in \Gamma_{5} \setminus \Gamma_4$ in \eqn{npl1biasexpand}, one can solve for the remaining $N_b^{(5)}-N_b^{(4)}$ basis elements using the rest of the degeneracy equations in \eqn{npl1degens}, and this will introduce the additional bias parameters that were not present at fourth order.  Since this is true for generic bias parameters $e_j^{(4)}(t)$, it is true in particular for the basis of descendants bias parameters $b_j^{(4)} ( t )$ in \eqn{bodexpand}.

\bibliography{matt_master_bib.bib}

\end{document}